\documentclass[conference,a4paper]{IEEEtran}
\usepackage{amsmath,amssymb,amsfonts}
\usepackage{algorithmic}
\usepackage{amsthm}
\usepackage{dsfont}
\usepackage{graphicx}
\usepackage{epsfig}
\usepackage{textcomp}
\usepackage{cite}
\usepackage{theoremref}
\def\BibTeX{{\rm B\kern-.05em{\sc i\kern-.025em b}\kern-.08em
    T\kern-.1667em\lower.7ex\hbox{E}\kern-.125emX}}
\usepackage[caption=false]{subfig}
\newcommand{\modphi}{\mbox{\large $\varphi$}}	
\newcommand{\rb}{\text{b}}										
\newcommand{\rc}{\text{c}}										
\newcommand{\re}{\mbox{\tiny ${\rm{e}}$}}			
\newcommand{\rf}{\mbox{\tiny ${\rm{f}}$}}			
\newcommand{\rFL}{\mbox{\tiny ${\rm{FL}}$}}			
\newcommand{\rL}{\mbox{\tiny ${\rm{L}}$}}			
\newcommand{\rPFL}{\text{\tiny{PFL}}} 				
\newcommand{\rR}{\mbox{\tiny ${\rm{R}}$}}			
\newcommand{\tT}{^{\text{T}}} 								
\DeclareMathOperator{\sgn}{sgn}								
\begin{document}
%
\title{Combined Sparse Regularization\\ for Nonlinear Adaptive Filters}
%
\author{\IEEEauthorblockN{Danilo~Comminiello\IEEEauthorrefmark{1}, Michele~Scarpiniti\IEEEauthorrefmark{1}, Simone~Scardapane\IEEEauthorrefmark{1}, Luis~A.~Azpicueta-Ruiz\IEEEauthorrefmark{2}, and~Aurelio~Uncini\IEEEauthorrefmark{1}}
\IEEEauthorblockA{\IEEEauthorrefmark{1}\textit{Dept. Information Eng., Electronics and Telecom. (DIET)}, \textit{Sapienza University of Rome, Italy}}
\IEEEauthorblockA{\IEEEauthorrefmark{2}\textit{Dept. Signal Theory and Communic.}, \textit{Universidad~Carlos~III~de~Madrid, Legan\'{e}s,~Spain}}
\IEEEauthorblockA{Corresponding author email: danilo.comminiello@uniroma1.it}}
%
\maketitle
%
%
\begin{abstract}
Nonlinear adaptive filters often show some sparse behavior due to the fact that not all the coefficients are equally useful for the modeling of any nonlinearity. Recently, a class of proportionate algorithms has been proposed for nonlinear filters to leverage sparsity of their coefficients. However, the choice of the norm penalty of the cost function may be not always appropriate depending on the problem. In this paper, we introduce an adaptive combined scheme based on a block-based approach involving two nonlinear filters with different regularization that allows to achieve always superior performance than individual rules. The proposed method is assessed in nonlinear system identification problems, showing its effectiveness in taking advantage of the online combined regularization.
\end{abstract}
%
%
%
%
%
\section{Introduction}
\label{sec:intro}
Very often, in estimating an impulse response that characterizes an unknown system, we may notice some kind of sparsity. Such sparse behavior of the system response is often due to the fact that most of its energy is contained in a small part of it \cite{Huang2006}. This means that a small part of the impulse response is characterized by large magnitude coefficients, while the rest shows negligible values. In system identification problems, this behavior can be exploited to improve the overall modeling performance by introducing a regularizing penalty in the cost function. In this sense, sparse regularization is widely employed in several fields of application, highlighting recently some examples found in underwater communication \cite{PanayirciTSP2016}, compressive sensing \cite{PantISCAS2017}, Alzheimer's disease diagnosis \cite{LeiTCYB2017}, wireless communications \cite{MasoodTSP2015}, adaptive beamforming \cite{ShiTSP2015,DeandradeTSP2015} and large-scale classification \cite{ScardapaneNEUCOM2017}, among others.

Sparsity may characterize not only linear systems but also nonlinear ones. In nonlinear modeling problems, which can be found in several real-world applications \cite{ComminielloBook2018}, nonlinear expansions or transformation may introduce a large number of coefficients, some of which can result to be useless for the purpose of nonlinear modeling. This often leads to overfitting phenomena and performance degradation \cite{ComminielloSIGPRO2017}. In order to tackle this problem and avoid any deterioration in performance, we may think to involve any methodology capable of selecting only the most significant nonlinear elements, i.e., those providing the best modeling performance. However, in real-world problems that require an online estimation, this kind of selection process is often not so trivial to be implemented. In particular, when a nonlinearity is nonstationary or depending on a time-varying signal, it is not possible to select \textit{a priori} the most significant nonlinear coefficients. In linear-in-the-parameters (LIP) nonlinear filters, one of the most effective approach to avoid overfitting is represented by adaptive algorithms involving sparse regularization \cite{ComminielloTASLP2014}. In the last years, several sparse nonlinear algorithms have been proposed, including kernel-based methods \cite{BazerqueSPM2013,LiuTNN2009} and polynomial methods \cite{KekatosTSP2011}, among others.

In this paper we focus on the class of LIP nonlinear filters known as functional link adaptive filters (FLAFs) \cite{ComminielloTASLP2014, ComminielloMLSP2016, ComminielloSIGPRO2017, ComminielloEUSIPCO2017}. Functional link models are widespread in the literature due to their flexibility and the large range of application in which they can be used \cite{SicuranzaTASL2011, ComminielloTASL2013, PatelTCASI2016, CariniICASSP2017}. They are basically characterized by a nonlinear transformation of the input by any nonlinear series expansion. Depending on the problem and on the nonlinearity to model, the number of functional links may be rather high, thus causing sparsity in the FLAF coefficient vector. However, this behavior can be exploited by imposing any sparsity constraint in the minimization problem. In that sense, proportionate regularization has been mainly considered to develop proportionate FLAFs (PFLAFs) achieving promising results \cite{ComminielloTASLP2014, ComminielloSIGPRO2017}. However, the possibility of using an $\ell_1$-norm constraint besides the classic proportionality can also be advantageous due to its desirable properties of inducing sparsity while keeping the convexity of the cost function \cite{TibshiraniJRSSB1996,ComminielloISCAS2018}. One of the most successfully used class of online adaptive algorithms showing an $\ell_1$ relaxation is based on zero-attracting proportionate adaptation \cite{ChenICASSP2009, ChenSPL2016, DasTASLP2016}.

In order to take always advantage of different sparse regularization rules, we propose the use of a combined nonlinear filtering scheme that permits to improve modeling performance when the \textit{a priori} knowledge about the nonlinearity to be modeled is limited \cite{ArenasSPM2016,ComminielloSIGPRO2013}. In particular, the proposed scheme involves an adaptive convex combination of two PFLAFs, involving one with a $\ell_1$-norm constraint. Moreover, we employ a block-based strategy that takes into account the cyclic nature of sparse functional links\cite{ComminielloSIGPRO2017}.

The paper is organized as follows. Section \ref{sec:flafapproach} briefly reviews the sparse functional link modeling, while two different sparse regularization rules are introduced in Section \ref{sec:pflaf}. In Section~\ref{sec:bbcomb}, we introduce the block-based combination scheme for the sparse FLAFs, and in Section \ref{sec:results} experimental results are shown. Concluding remarks are drawn in Section \ref{sec:conclusion}.
%
%
%
%
%
\section{Nonlinear Modeling with Sparse FLAF}
\label{sec:flafapproach}
%
\subsection{A Review of the Functional Link Adaptive Filter}
\label{subs:flafmodel}
The functional link adaptive filter (FLAF) \cite{ComminielloTASL2013} is a LIP nonlinear filter that expands a linear input signal ${\mathbf{x}}_{\rL,n}  \in \mathbb{R}^M  = \left[ {\begin{array}{*{20}c} {x\left[n\right]} & {x\left[n-1\right]} & \ldots & {x\left[n-M+1\right]}  \\ \end{array}} \right]\tT$, being $M$ the length of the regression vector, in order to filter it in a higher dimensional space. The transformation of the input signal is performed by applying a nonlinear expansion series in order to produce the nonlinear signal. The chosen functions of the expansion series represent the set of functional links $\Phi = \left\{\modphi_{\scriptscriptstyle 0}\left(\cdot\right),\ldots,\modphi_{\scriptscriptstyle Q_{\rf}-1}\left(\cdot\right)\right\}$, where $Q_{\rf}$ is the number of the chosen functional links. One of the most popular choice for populating the functional link set is to use the trigonometric series expansion, which can be described as:
\begin{equation}
	\modphi_{\scriptscriptstyle j}\left(x\left[n-i\right]\right) = \left\{ {\begin{array}{*{20}c}
   {\sin \left( {p\pi x\left[ {n - i} \right]} \right),} \hfill & {j = 2p - 2} \hfill  \\
   {\cos \left( {p\pi x\left[ {n - i} \right]} \right),} \hfill & {j = 2p - 1} \hfill  \\
	\end{array}} \right.
	\label{eq:trigoser}
\end{equation}

\noindent for $i = 0, \ldots, M-1$. In \eqref{eq:trigoser}, $p = 1,\ldots,P$ is the expansion index, where $P$ is the expansion order, and $j=0,\ldots,Q_{\rf}-1$ is the functional link index. For the trigonometric memoryless expansion, the overall number of functional links contained in the set $\Phi$ is equal to $Q_{\rf} = 2P$.
%

The overall expanded vector resulting from the application of the functional link set to the input signal is denoted as $\mathbf{g}_n \in \mathbb{R}^{M_{\re}} = \left[\begin{array}{*{20}c} {g_{0}\left[n\right]} & {g_{1}\left[n\right]} & \ldots & {g_{M_{\re}-1}\left[n\right]} \end{array}\right]\tT$, where $M_{\re} \geq M$ is the length of the expanded vector. Then, the signal $\mathbf{g}_n$, which represents the input in a higher dimensional space, can be processed by any adaptive filter to achieve the system output.

The functional link set $\Phi$ may contain both linear and nonlinear functions. However, we assume that all the functional links are nonlinear \cite{ComminielloTASL2013}, such that the resulting expanded vector $\mathbf{g}_n$ is completely composed of nonlinear elements. This allows to increase the flexibility of the adaptive scheme in those problems which require the modeling of linear and nonlinear components, since we can devote one adaptive filter $\mathbf{w}_{\rL,n} \in \mathbb{R} ^M  = \left[ {{\begin{array}{*{20}c} {w_{\rL,0}\left[ n \right]} &  \ldots  \\ \end{array}}}\right.$ $\left. {{\begin{array}{*{20}c} {w_{\rL,M-1}\left[ {n} \right]} \\ \end{array}}} \right]\tT$ for the non-expanded linear input ${\mathbf{x}}_{\rL,n}$ and another adaptive filter $\mathbf{w}_{\rFL,n} \in \mathbb{R} ^{M_{\re}}  = \left[ {{\begin{array}{*{20}c} {w_{\rFL,0}\left[ n \right]} &  \ldots  \\ \end{array}}}\right.$ $\left. {{\begin{array}{*{20}c} {w_{\rFL,M_{\re}-1}\left[ {n} \right]} \\ \end{array}}} \right]\tT$ for $\mathbf{g}_n$, enabling us to choose different learning algorithms with different capabilities for each adaptive filter. This method is detailed in \cite{ComminielloTASL2013}.
%
\subsection{Sparsity in FLAF}
\label{subs:sparsityFL}
The concept of sparsity in a linear filter has been dealt with quite extensively in the literature. However, a major attention is required for a LIP nonlinear filter, as in the case of $\mathbf{w}_{\rFL,n}$.

In order to understand how sparsity behaviors occur in FLAFs, it is sufficient to analyze the energy of the coefficient vector $\mathbf{w}_{\rFL,n}$ at steady state \cite{ComminielloSIGPRO2017}, i.e., for $n \to \infty$. As a result, the early functional links of the set $\Phi$, which have small values of the expansion order (i.e., $p$ close to $1$), generate the most significant nonlinear elements for the purpose of the nonlinear modeling. On the other hand, the remaining functional links of $\Phi$ produce only minor, or even negligible, variations in the modeling results. This is physically motivated by the fact that high-order functional links (i.e., with $p \to P$) aim at modeling those nonlinear components that not always appear in a nonlinear distortion (e.g., high-order harmonics), thus resulting in a slight improvement that might not always occur. In terms of energy, it is possible to describe the sparsity behavior in functional links as an exponential decay from early to late elements, in which the larger the expansion order the longer the tail. Such behavior occurs for each input sample that is expanded by the functional link set $\Phi$, therefore, overall, we may have a periodic sparse behavior for $\mathbf{w}_{\rFL,\infty}$ \cite{ComminielloSIGPRO2017}.
%
%
%
%
%
\section{Sparse Functional Link Adaptive Filters}
\label{sec:pflaf}
In this section, we derive two sparse FLAFs, $\mathbf{w}_{\rFL1,n}$ and $\mathbf{w}_{\rFL2,n}$, respectively using an $\ell_1$-norm proportionate and a classic proportionate regularization.
%
\subsection{Derivation of the $\ell_1$-Regularized PFLAF}
\label{subs:l1pflaf}
In order to derive the optimization algorithms for the $\ell_1$ sparse functional links, we can express the $\ell_1$-constrained optimization problem, considering the least-perturbation property and the natural gradient adaptation, as suggested in \cite{ComminielloTASLP2014}, thus:
\begin{gather} \label{eq:localproblem}
		\arg \mathop {\min }\limits_{\mathbf{w}_{\rFL1,n}} \left\|\mathbf{w}_{\rFL1,n} - \mathbf{w}_{\rFL1,n-1}\right\|_{\mathbf{Q}_{1,n}^{-1}}^2 + \gamma\left\|\mathbf{Q}_{1,n}^{-1}\mathbf{w}_{\rFL1,n}\right\|_1 \\
		\text{s.t.} \ \ \ \varepsilon_1\left[n\right] = 0 \nonumber
\end{gather}

\noindent where $\mathbf{Q}_{1,n}^{-1}$ is a distance correction matrix with respect to the Euclidean metric, $\gamma$ is a very small constant and the constraint $\varepsilon_1\left[n\right] = 0$ can be derived from the \textit{a posteriori} output estimation error signal $\varepsilon_1\left[n\right] = d\left[n\right] - \mathbf{x}_{\rL,n}\tT\mathbf{w}_{\rL,n} - \mathbf{g}_{n}\tT\mathbf{w}_{\rFL1,n}$. The problem in \eqref{eq:localproblem} involves an $\ell_1$-norm penalty term that aims at scaling the coefficient vector $\mathbf{w}_{\rFL1,n}$ by the distance correction matrix $\mathbf{Q}_{1,n}^{-1}$ and thus regularizing the solution by mainly exploiting inactive coefficients.

The solution to the problem \eqref{eq:localproblem}, with respect to $\mathbf{w}_{\rFL1,n}$, can be expressed as:
\begin{equation}
	\begin{split}
		\mathbf{w}_{\rFL1,n} &= \mathbf{w}_{\rFL1,n-1} + \mu_{\rFL} \frac{\mathbf{Q}_{1,n}\mathbf{g}_n e_{\rFL1}\left[n\right]}{\mathbf{g}_n\tT\mathbf{Q}_{1,n}\mathbf{g}_n + \delta}\\
		&- \gamma_{\rR}\left[n\right]\frac{\sgn\left(\mathbf{w}_{\rFL1,n-1}\right)}{1 + \epsilon\left|\mathbf{w}_{\rFL1,n-1}\right|},
	\end{split}
	\label{eq:vssrzaupdate}
\end{equation}

\noindent where $e_{\rFL1}\left[n\right] = d\left[n\right] - \mathbf{x}_{\rL,n}\tT\mathbf{w}_{\rL,n-1} - \mathbf{g}_{n}\tT\mathbf{w}_{\rFL1,n-1} = d\left[n\right] - y_{\rL}\left[n\right] - y_{\rFL1}\left[n\right]$, $\delta$ is a regularization factor, $\mu_{\rFL1}$ is a step-size parameter, and $\sgn\left(\cdot\right)$ is intended as an element-wise sign function defined for the $i$-th entry of $\mathbf{w}_{\rFL1,n}$ as:
\begin{equation}
	\sgn\left( {w_{\rFL1,i}\left[n\right]} \right) = \left\{ {\begin{array}{*{20}c}
   {{{w_{\rFL1,i} \left[ n \right]} \mathord{\left/
 {\vphantom {{w_{\rFL1,i} \left[ n \right]} {\left| {w_{\rFL1,i} \left[ n \right]} \right|,}}} \right.
 \kern-\nulldelimiterspace} {\left| {w_{\rFL1,i} \left[ n \right]} \right|,}}} & {w_{\rFL1,i} \left[ n \right] \ne 0}  \\
   {0,} & {w_{\rFL1,i} \left[ n \right] = 0}  \\
\end{array}} \right.
	\label{eq:signdef}
\end{equation}

\noindent Equation \eqref{eq:vssrzaupdate} involves some considerations. First, we assumed that most of the weights do not change their sign at each iteration, especially those weights that are inactive, i.e, that are close to zero at steady state. Therefore, the term $\sgn\left(\mathbf{w}_{\rFL1,n}\right)$, which contains the unknown new update, can be approximated by $\sgn\left(\mathbf{w}_{\rFL1,n-1}\right)$ without affecting the convergence behavior \cite{ChenICASSP2009, DasTASLP2016}. 

Moreover, with respect to original PFLAFs \cite{ComminielloEUSIPCO2017, ComminielloSIGPRO2017}, the FLAF based on the $\ell_1$ penalty involves a third additional term, the ``zero attractor'', which shrinks the inactive weights of $\mathbf{w}_{\rFL1,n-1}$ to zero. However, such term may lose its effectiveness as the sparsity of a system decreases, i.e., the number of active coefficients increases. To overcome this limitations, the $\ell_1$ regularization term $\left\|\mathbf{Q}_{1,n}^{-1}\mathbf{w}_{n}\right\|_1$ in \eqref{eq:localproblem} has been replaced by a log-sum penalty term $\left\|\mathbf{Q}_{1,n}^{-1}\log\left(1+\epsilon\mathbf{w}_{\rFL1,n}\right)\right\|_1$, where $\epsilon$ is a small constant \cite{CandesJFAA2008, ChenICASSP2009}.

Finally, in \eqref{eq:vssrzaupdate}, $\gamma_{\rR}\left[n\right] = \epsilon \gamma \mu_{\rR}\left[n\right]$, where $\mu_{\rR}\left[n\right]$ is a nonparametric variable step size (VSS) \cite{PaleologuTASL2008, ComminielloISCAS2018}, defined as:
\begin{equation}
	\mu_{\rR}\left[n\right] = \left|1 - \frac{\sqrt{\left|\widehat{\sigma}_d^2\left[n\right] - \widehat{\sigma}_{y_{\rL}}^2\left[n\right]- \widehat{\sigma}_{y_{\rFL1}}^2\left[n\right]\right|}}{\widehat{\sigma}_{e_{\rFL1}}^2\left[n\right] + \xi}\right|.
	\label{eq:vss}
\end{equation}

\noindent In \eqref{eq:vss}, the general parameter $\widehat{\sigma}_{\theta}^2\left[n\right]$ represents the power estimate of the sequence $\theta\left[n\right]$, being $\theta = \left\{d, y, e\right\}$, and it can be computed as $\widehat{\sigma}_{\theta}^2\left[n\right] = \beta\widehat{\sigma}_{\theta}^2\left[n-1\right] + \left(1 - \beta\right)\theta^2\left[n\right]$ where $\beta \to 1$ is a forgetting factor. $\xi$ is a small positive constant that avoids divisions by zero.

The update equation \eqref{eq:vssrzaupdate} defines the VSS reweighted zero attractor FLAF. The last term of \eqref{eq:vssrzaupdate}, in which the division is performed element-wise, is the reweighted zero attractor, whose aim is to shrink to zero those coefficients whose magnitude is comparable to $1/\epsilon$, thus preserving the most active coefficients. This usually results in an improvement in terms of convergence rate and steady state performance.
%
%
\subsection{Derivation of the Proportionate FLAF}
\label{subs:l2pflaf}
The second sparse FLAF can be derived by formalizing the following optimization problem:
\begin{align} \label{eq:localproblem2}
		&\arg \mathop {\min }\limits_{\mathbf{w}_{\rFL2,n}} \left\|\mathbf{w}_{\rFL2,n} - \mathbf{w}_{\rFL2,n-1}\right\|_{\mathbf{Q}_{2,n}^{-1}}^2 \\
		\text{s.t.} \ \varepsilon_2\left[n\right] = &\left(1 - \frac{\mu_{\rL}\left\|\mathbf{x}_{\rL,n}\right\|_2^2}{\left\|\mathbf{x}_{\rL,n}\right\|_2^2 + \delta_{\rL}} - \frac{\mu_{\rFL2}\left\|\mathbf{g}_{n}\right\|_{\mathbf{Q}_{2,n}}^2}{\left\|\mathbf{g}_{n}\right\|_{\mathbf{Q}_{2,n}}^2 + \delta_{\rPFL}}\right) e_{\rFL2}\left[n\right] \nonumber
\end{align}

\noindent where $\mu_{\rL}$ and $\mu_{\rFL2}$ are step-size parameters, $\delta_{\rL}$ and $\delta_{\rPFL2}$ are regularization factors. The resulting solution, whose complete derivation can be found in \cite{ComminielloTASLP2014}, can be written as:
\begin{equation}
	\mathbf{w}_{\rFL2,n} = \mathbf{w}_{\rFL2,n-1} + \mu_{\rFL2}\frac{\mathbf{Q}_{2,n}\mathbf{g}_n}{\mathbf{g}_n{\tT}\mathbf{Q}_{2,n}\mathbf{g}_n + \delta_{\rPFL}}e_{\rFL2}\left[n\right].
	\label{eq:PFLAFupdate}
\end{equation}
%
\subsection{Choice of the weighting matrix}
\label{subs:wmat}
The matrix $\mathbf{Q}_{j,}n$, with $j=1,2$, in \eqref{eq:vssrzaupdate} and \eqref{eq:PFLAFupdate}, also known as proportionate matrix, aims at weighting the coefficients of $\mathbf{w}_{\rFL j,n}$ proportionally to the contribution they provide to the modeling. $\mathbf{Q}_{j,n}$ can be chosen as a diagonal matrix $\mathbf{Q}_{j,n} \in \mathbb{R}^{M_{\re}} = {\text{diag}}\left\{ {\begin{array}{*{20}c} {q_{j,0}\left[n\right]} & \ldots & {q_{j,M_{\re}-1}\left[n\right]} \\ \end{array}} \right\}$, whose diagonal elements are derived according to the coefficients at the time instant $n-1$.

Several choices can be made for deriving the diagonal elements of $\mathbf{Q}_{j,n}$. We choose the same derivation for both the sparse FLAFs according to \cite{BenestyICASSP2002,ComminielloTASLP2014}:
\begin{equation}
	q_{j,k}\left[n\right] = \frac{1 - \alpha}{2M_{\re}} + \left(1 + \alpha\right) \frac{\left|w_{\rFL j,k}\left[n-1\right]\right|}{\xi + 2 \sum_{i=a}^{b}w_{\rFL j,k\left[n-1\right]},}
	\label{eq:qcoeff}
\end{equation}

\noindent with $k = 0,\ldots,M_{\re}-1$ and $\xi$ being a small positive value. In \eqref{eq:qcoeff}, the proportionality factor $-1 \leq \alpha \leq 1$ assumes a value close to $1$ when a high degree of sparseness is expected, while, on the contrary, a low degree is expected when the proportionality factors are close to $-1$.
%
%
%
%
%
\section{Combined Sparse Regularization Scheme}
\label{sec:bbcomb}
\begin{figure}[t]
	\centering
	\includegraphics[width=0.45\textwidth,keepaspectratio]{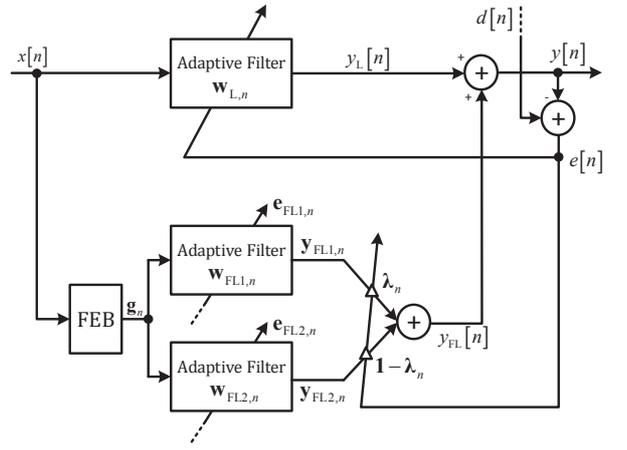}
	\caption{The proposed combined scheme where the filter block outputs are combined by the mixing parameter vectors.}
	\label{fig:bbfpsfaf}
\end{figure}

In order to take advantage of the two sparse FLAFs, we adopt an adaptive combined architecture characterized by a linear branch and, in parallel, a convex combination of the FLAFs, as represented in Fig.~\ref{fig:bbfpsfaf}.

The output of the linear branch is performed as a classic linear filtering, while the nonlinear branch involves the generation of the expanded signal $\mathbf{g}_n$, as previously described, which is then fed into both the adaptive filters, thus generating the individual outputs and errors, $y_{\rFL j}\left[n\right] = \mathbf{g}_n\tT\mathbf{w}_{\rFL j,n-1}$ for $j=1,2$. The overall output of the nonlinear branch is achieved by combining convexly the individual filter outputs in a block-based fashion, as described in \cite{ComminielloSIGPRO2017}. 

The block-based combination takes advantage of the periodic nature of the sparse energy behavior of functional links, as detailed in \cite{ComminielloSIGPRO2017}. Each group of expanded samples may be characterized by a similar energy decay to that of the adjacent groups. Based on this property, we may think to divide the coefficient vectors $\mathbf{w}_{\rFL 1,n}$ and $\mathbf{w}_{\rFL 2,n}$ into $L$ blocks, each one consisting of $M_{\rb} = Q_{\rf}/L$ coefficients. Therefore, the output of the block-based combination can be written as:
\begin{equation}
	\begin{split}
		y_{\rFL}\left[n\right] &= \sum_{i = 0}^{M-1} \left( \sum_{l=1}^{L} \lambda_{l}\left[n\right] \mathbf{g}_{n}^{\left(iL+l\right){\text{T}}} \mathbf{w}_{\rFL1,n-1}^{\left(iL+l\right)} \right.\\
		&+ \left.\left(1 - \lambda_{l}\left[n\right]\right) \mathbf{g}_{n}^{\left(iL+l\right){\text{T}}} \mathbf{w}_{\rFL2,n-1}^{\left(iL+l\right)} \right)
	\end{split}
	\label{eq:newbbout}
\end{equation}

\noindent where the index $i$ denotes the input signal sample that is expanded by \eqref{eq:trigoser}, $l$ represents the block index and $\lambda_{l}\left[n\right]$ is the mixing parameter associated to the $l$-th block. The $l$-th adaptive mixing parameter $\lambda_{l}\left[n\right]$, with $l=1,\ldots,L$, in \eqref{eq:newbbout} balances the combination between the $l$-th blocks of the two filters $\mathbf{w}_{\rFL j}\left[n\right]$ ($j = 1, 2$), giving more importance to the best performing filter block \cite{ArenasSPM2016}. Such awareness is obtained according to a mean-square error minimization. In particular, the adaptation of $\lambda_{l}\left[n\right]$ is performed by using an auxiliary adaptive parameter for each block $a_{l}\left[n\right]$, which is related to $\lambda_{l}\left[n\right]$ by means of a sigmoid function that keeps the mixing parameter in the range $\left[0,1\right]$ \cite{ArenasSPM2016,ComminielloSIGPRO2013}:
\begin{equation}
	\lambda_{l}\left[n\right] = \eta\left(\frac{1}{1 + e^{-a_{l}\left[n\right]}} - \theta\right),
	\label{eq:relationlambdaa}
\end{equation}

\noindent where $\theta = 1/\left(1 + e^4\right)$ and $\eta = 1/\left(1-2\theta\right)$. The update rule for the auxiliary parameter for the $l$-th block is:
\begin{equation}
	\begin{split}
		a_{l}\left[n\right] = &a_{l}\left[n-1\right] + \frac{\mu_{\rc}}{\eta r_{l}\left[n-1\right]} e\left[n\right] \Delta y_{\rFL,l}\left[n\right]\\
		&\cdot \left(\lambda_{l}\left[n\right] + \theta\eta\right) \left(\eta-\theta\eta-\lambda_{l}\left[n\right]\right)
	\end{split}
	\label{eq:updateap}
\end{equation}
\noindent for $l = 1,\ldots,L$, where
\begin{equation}
	\Delta y_{\rFL,l}\left[n\right] = \sum_{i=1}^{M} \mathbf{g}_n^{\left(i,l\right){\text{T}}} \left(\mathbf{w}_{\rFL1,n-1}^{\left(i,l\right)} - \mathbf{w}_{\rFL2,n-1}^{\left(i,l\right)}\right),
	\label{eq:delta1FL}
\end{equation}

\noindent where the superscript $\left(i,l\right)$ denotes the $l$-th block related to the $i$-th entry. In \eqref{eq:updateap}, $\mu_{\rc}$ is the step-size parameter of the adaptive combination, $r_{l}\left[n\right] = \beta_r r_{l}\left[n-1\right] + \left(1-\beta_r \right) \Delta y_{\rFL,l}^2\left[n\right]$ is the estimated power of $\Delta y_{\rFL,l}\left[n\right]$ that permits a normalized adaptation of $a_{l}\left[n\right]$, and $0 \ll \beta_r < 1$ is a smoothing factor. It is worth noting that eq.~\eqref{eq:delta1FL} takes into account the periodic nature of the functional link expansions.

Once achieved the filtering outputs $y_{\rL}\left[n\right]$ and $y_{\rFL}\left[n\right]$, it is possible to derive the error signals used for the adaptation of the filters on the nonlinear branch $e_{\rFL j}\left[n\right] = d\left[n\right] - \left(y_{\rL}\left[n\right] + y_{\rFL j}\left[n\right]\right)$, and also the overall error signal $e\left[n\right] = d\left[n\right] - \left(y_{\rL}\left[n\right] + y_{\rFL}\left[n\right]\right)$, which is used for the adaptation of both the weight vector $\mathbf{w}_{\rL,n}$ and the $L$ auxiliary parameters in \eqref{eq:updateap}.
%
%
%
%
%
\section{Simulation Results}
\label{sec:results}
In order to assess the proposed scheme we consider a nonlinear system identification problem, in which a linear system is preceded by a nonlinear one that applies a soft-clipping nonlinearity to the input signal \cite{ComminielloTASLP2014}:
\begin{equation}
	\overline{y}\left[n\right] = \left\{ {\begin{array}{*{20}c}
   {2 x\left[ n \right]/{3\zeta}} &, & {0 \le \left|x\left[ n \right]\right| \le \zeta }  \\
   {\frac{{3 - \left( {2 - \left|x\left[ n \right]\right|/\zeta} \right)^2 }}{3}\sgn\left( {x\left[ n \right]} \right)} &, & {\zeta  \le \left|x\left[ n \right]\right| \le 2\zeta }  \\
   {\sgn\left( {x\left[ n \right]} \right)} &, & {2\zeta  \le \left|x\left[ n \right]\right| \le 1}  \\
\end{array}} \right.
	\label{eq:clip}
\end{equation}

\noindent where $0 < \zeta \leq 0.5$ is a nonlinearity threshold. The linear system is formed by $M = 15$ independent random values between $-1$ and $1$. White Gaussian noise $v\left[n\right]$ is added at output of the nonlinear system with $30$ dB of signal-to-noise ratio (SNR). The input signal $x\left[n\right]$ is a colored Gaussian noise with length $L = 40000$. In order to evaluate performance we use the \textit{excess mean-square error} (EMSE) in dB ${\rm{EMSE}}\left[n\right] = 10 \log _{10} \left({\rm{E}}\left\{\left(e\left[n\right] - v\left[n\right]\right)^2\right\}\right)$, which is averaged over $1000$ runs with respect to input and noise. The parameter setting for both the FLAFs is: $\mu_{\rFL} = 0.1$, $\delta = 10^{-3}$, $P = 20$, $\gamma = 10^{-5}$, $\epsilon = 10^{-2}$, $\beta = 0.99$, $\alpha = 0$. The filter $\mathbf{w}_{\rL,n}$ is adapted by an NLMS algorithm with a step size $\mu_{\rL} = 0.1$ and a regularization parameter also set to $\delta = 10^{-3}$. For the block-based combination we set $\mu_{\rc} = 0.1$, $\beta_r = 0.9$ and initial values $a_l\left[0\right] = 0$ and $r_l\left[0\right] = 1$.
\begin{figure}[t]
	\centering
	\includegraphics[width=0.45\textwidth,keepaspectratio]{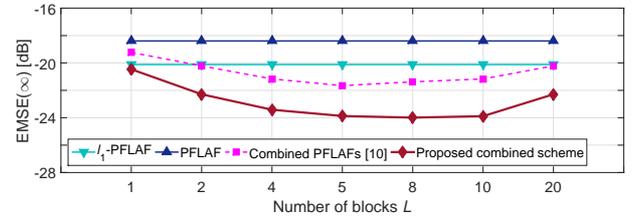}
	\caption{Steady-state EMSE of the proposed method on varying the number of blocks in the combination.}
	\label{fig:exp1_ssemse}
\end{figure}
\begin{figure}[t]
	\centering
	\subfloat[]{\includegraphics[width=0.45\textwidth,keepaspectratio]{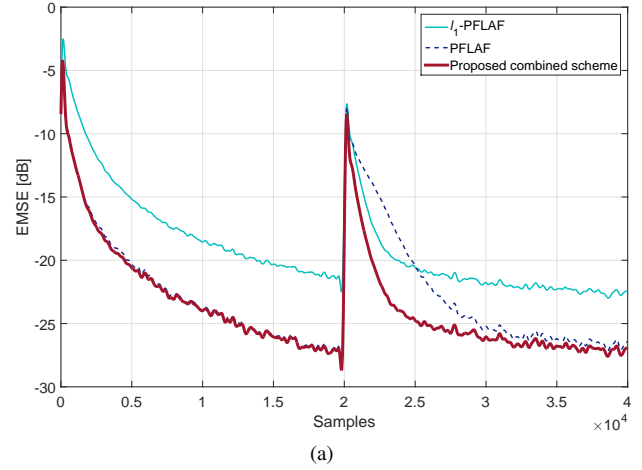}\label{fig:exp1_emse_behavior}}
	\vfill
	\subfloat[]{\includegraphics[width=0.45\textwidth,keepaspectratio]{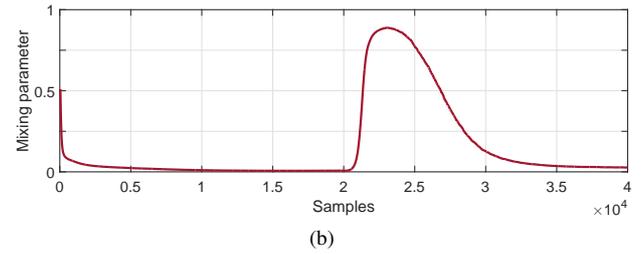}\label{fig:exp1_emse_mixpar}}
	\caption{(a) Convergence behavior in terms of the EMSE of the proposed combined FLAF with $L=8$ and (b) related mixing parameter evolution of the first block.}
	\label{fig:exp1_emse}
\end{figure}

We evaluate the steady-state performance in the case of strong nonlinearity level ($\zeta = 0.03$). Results are depicted in Fig.~\ref{fig:exp1_ssemse}, where the values of the steady-state EMSE are shown on varying the number of blocks from $L=1$, which implies the adaptation of a full block with size $M_{\rb} = Q_{\rf} = 40$, to $L=20$, which implies the adaptation of blocks with size $M_{\rb} = Q_{\rf}/L = 2$. We compare the performance of the proposed scheme with that of individual sparse FLAFs involving $\ell_1$ and proportionate regularization separately, and the performance of the same scheme involving the combination of two PFLAFs (i.e., the algorithm cPSFLAF\#1 in \cite{ComminielloSIGPRO2017}). Results show that the use of a blockwise combination (i.e., $L>1$) always produces better results with respect to the full-block combination. The best performance is achieved with $L=8$, which gains about $4$ dB over the same algorithm with $L=1$. Reducing too much the block size leads to a performance decrease, due to gradient noise in the adaptation of the mixing parameters, and also to a computational cost increase.

We also assess the tracking performance by setting two different nonlinearity levels: $\zeta = 0.08$ for the first half of the experiment and $\zeta = 0.05$ (i.e., stronger nonlinearity) for the second half. Results in Fig.~\ref{fig:exp1_emse}\subref{fig:exp1_emse_behavior}, show the improvement of the proposed method (with $L=8$) over the other ones. In particular, it takes advantage of the steady-state performance of the sparse PFLAF, while it exploits the faster convergence rate provided by the $\ell_1$-sparse PFLAF when nonlinearity changes. This result is also confirmed by the mixing parameter evolution, depicted in Fig.~\ref{fig:exp1_emse}\subref{fig:exp1_emse_mixpar} for the first block.
%
%
%
%
\section{Conclusion}
\label{sec:conclusion}
In this paper, a combined sparse regularization scheme for LIP nonlinear filters has been developed based on an $\ell_1$-norm proportionate and a classic proportionate sparsity. The proposed method involves a blockwise adaptive combinations of sparse FLAFs having different adaptation rules aiming at improving the overall modeling performance. Experimental results have shown the effectiveness and robustness of our proposal in a time-varying nonlinear system identification problem.
%
%
%
%
%
\section*{Acknowledgment}
The work of M. Scarpiniti, S. Scardapane and A. Uncini is partially supported by the Italian National Project ``GAUChO - A Green Adaptive Fog Computing and Networking Architecture'', under grant number 2015YPXH4W. 

The work of L.~A.~Azpicueta-Ruiz has been partly supported by MINECO Projects TEC2014-52289-R and TEC2017-83838-R.
%
%
%
%
%
\bibliographystyle{IEEEtran}
\bibliography{EUSIPCO2018_SFL_v5_arXiv}
%
%
\end{document}